\documentclass[journal=jacsat,manuscript=article]{achemso}

\usepackage[version=3]{mhchem} % Formula subscripts using \ce{}

\author{Vasileios Fotopoulos}
\email{vasileios.fotis.19@ucl.ac.uk}
\affiliation[University College London]
{Department of Physics and Astronomy, University College London, Gower Street, London WC1E 6BT, United Kingdom}
\author{Corey S. O'Hern}
\affiliation[Yale University1]{Department of Mechanical Engineering $\&$ Materials Science, Yale University, New Haven, 06520, CT, USA}
\alsoaffiliation[Yale University]{Department of Physics, Yale University, New Haven, 06520, CT, USA}
\alsoaffiliation[Yale University]{Department of Applied Physics, Yale University, New Haven, 06520, CT, USA}
\author{Mark D. Shattuck}
\affiliation[The City College of the City University of New York]
{Benjamin Levich Institute and Physics Department, The City College of the City University of New York, New York, NY, USA}
\author{Alexander L. Shluger}
\affiliation[University College London]
{Department of Physics and Astronomy, University College London, Gower Street, London WC1E 6BT, United Kingdom}
\alsoaffiliation[Tohoku University]{WPI-Advanced Institute for Materials Research (WPI-AIMR), Tohoku University, 2–1-1 Katahira, Aoba-ku,
Sendai 980-8577, Japan}

\title[An \textsf{achemso} demo]
  {Modeling the effects of varying Ti concentration on the mechanical properties of Cu-Ti alloys}

\abbreviations{IR,NMR,UV}

\begin{document}

\begin{abstract}
 The mechanical properties of Cu-Ti alloys have been characterized extensively through experimental studies. However, a detailed understanding of why the strength of Cu increases after a small fraction of Ti atoms is added to the alloy is still missing. In this work, we address this question using density functional theory (DFT) and molecular dynamics (MD) simulations with modified embedded atom method (MEAM) interatomic potentials. First, we performed calculations of uniaxial tension deformations of small bicrystalline Cu cells using DFT static simulations. We then carried out uniaxial tension deformations on much larger bicrystalline and polycrystalline Cu cells using MEAM MD simulations. In bicrystalline Cu, the inclusion of Ti increases the grain boundary separation energy and maximum tensile stress. The DFT calculations demonstrate that the increase in tensile stress can be attributed to an increase in the local charge density arising from Ti. MEAM simulations in larger bicrystalline systems have shown that increasing the Ti concentration decreases the density of stacking faults. This observation is enhanced in polycrystalline Cu, where the addition of Ti atoms, even at concentrations as low as 1.5 at.\%, increases the yield strength and elastic modulus of the material compared to pure Cu. Under uniaxial tensile loading, the addition of small amounts of Ti hinders the formation of partial Shockley dislocations in the grain boundaries of Cu, leading to reduced local deformation. These results shed light on the role of Ti in determining the mechanical properties of polycrystalline Cu and will enable the engineering of grain boundaries and the inclusion of Ti to improve degradation resistance.
\end{abstract}

\section{\label{sec:level1}Introduction}

Numerous experimental studies have shown that binary alloys made from Cu have significantly improved mechanical properties compared to pure Cu~\cite{an2011effects,chad2020stable}. For instance, Cu-Be alloys have been widely used in numerous applications~\cite{wang2022comparison,mazar2012characterization}. However, Be is highly toxic even in small amounts, and thus there is significant interest in identifying other Cu alloys with similar advantageous properties. In particular, Cu-Ti alloys, used in a wide range of applications~\cite{semboshi2014age} are promising since they possess high strength and electrical conductivity, as well as improved corrosion
resistance~\cite{datta1976structure,laughlin1975spinodal,nagarjuna1995effects}. Furthermore, Cu-Ti alloys have an increased mechanical strength over pure Cu even at concentrations less than 5 atomic (at.) \% Ti~\cite{eze2018effect,ramesh2020investigation}.

A detailed understanding of the atomistic mechanisms that improve the strength of Cu through the addition of small amounts of Ti can be achieved by using theoretical and computational modeling. Kohn-Sham~\cite{hohenberg1964inhomogeneous,kohn1965self} density functional theory (DFT) and other quantum mechanics-based calculation methods can provide highly accurate electronic structure measurements for alloys~\cite{yang2022simulating,alidoust2020density}. Several theoretical studies highlight the strengthening effect of metallic solutes when they are introduced into the grain boundaries (GBs) of metals, such as Cu~\cite{razumovskiy2018solute}, Ni~\cite{bentria2019toward}, V~\cite{wu2016effect} and Au~\cite{scheiber2021segregation}. However, these calculations are limited to systems with only $\sim 100$ atoms due to their computational cost~\cite{fabian2022linear}. However, understanding the structural and mechanical properties of polycrystalline alloys requires much larger system sizes than $\sim 100$ atoms~\cite{derlet2007million}. Large-scale molecular dynamics (MD) simulations using embedded atom method (EAM) potentials can reproduce the structural and mechanical properties of many alloys~\cite{daw1993embedded}. The modified EAM (MEAM) proposed by Baskes, \textit{et al.}~\cite{baskes1997determination} was developed to extend EAM interatomic potentials to alloys with strong angular bonding. MD simulations using MEAM potentials have been carried out to understand important properties of alloys, such as ductile versus brittle mechanical response~\cite{kang2007brittle}, structure-property relationships~\cite{gates2005computational}, dislocation dynamics~\cite{martinez2008atomistically}, and fracture mechanics~\cite{potirniche2006molecular} in FCC metals and alloys. However, MEAM potentials have been developed for only a small fraction of alloys~\cite{jelinek2012modified,kim2008modified}.

Several previous studies have investigated the effect of metallic dopants on the mechanical properties of Cu and similar metals using either DFT~\cite{huang2018uncovering} or semi-emprical MD simulations. However, identifying the origins of the experimentally-observed improvements in the mechanical properties of polycrystalline Cu-Ti compared to those of pure Cu requires understanding the role of topological defects, such as dislocations. In this study, we investigate how the addition of Ti atoms affects the mechanical properties of bicrystalline and polycrystalline Cu using {\it both} DFT calculations and MEAM MD simulations. First, we use DFT calculations involving relatively small periodic cells to assess the accuracy of a recent MEAM potential that was fitted for Cu-Ti~\cite{miraz2020development}. We find that the MEAM potential accurately predicts the most energetically favorable segregation sites of Ti at the Cu grain boundaries. Using DFT calculations of CuTi alloys undergoing uniaxial tension deformations, we show that the addition of Ti increases charge localization and the separation energy of Cu GBs, which indicates the formation of covalent bonding between Ti and its neighboring Cu atoms. We then carry out MEAM MD simulations of bicrystalline Cu systems with Ti impurities undergoing uniaxial tension deformation. We show that, similar to the DFT results, the presence of Ti increases the yield strength of bicrystalline Cu by hindering the formation of stacking faults. We then conduct MEAM MD simulations of polycrystalline Cu-Ti undergoing uniaxial tension deformations. We find that the addition of even a small amount of Ti increases the yield strength and Young's modulus of the Cu polycrystals. The increases in the yield strength and elastic modulus of the polycrystalline Cu are caused by the fact that Ti hinders the emission of Shockley dislocations from the GBs during the tensile deformation. This strengthening increases with the concentration of Ti atoms at the GBs. In light of recent experimental studies that developed advanced materials with controllable interfaces~\cite{zhou2023atomic}, our findings highlight the potential of decorating GBs in polycrystalline Cu with Ti atoms to improve its resistance to degradation.

\section{\label{sec:level1}Methods}

\subsection{\label{sec:level3}First-Principles Calculations}

The DFT calculations were carried out using the Vienna Ab Initio Simulation Package (VASP)~\cite{kresse1993ab, kresse1996efficient,kresse1996efficiency} with the Perdew–Burke–Ernzerhof (PBE) GGA exchange-correlation functional~\cite{perdew1996generalized} in 76- and 108-atom periodic cells. (Additional details concerning the DFT calculations can be found in Appendix A.) In line with previous studies of Cu~\cite{ganchenkova2014effects}, a mixture of the Davidson~\cite{er1975iterativecalculationof} and RMM-DIIS~\cite{pulay1980convergence,wood1985new} algorithms is used to minimize the energy. Relevant details about the functional form of the ground state energy (\begin{large}$\mathcal{E}$\end{large}) obtained via DFT/PBE can be found in the literature~\cite{perdew1996generalized}. Two $\Sigma5$ twin boundary symmetries were examined,  (210)[100] and (012)[100], using different cell sizes.  These GBs are among the lowest energy reported in Cu~\cite{wu2016first}. The GB simulation cells are periodically translated in the $x$-, $y$-, and $z$-directions. In the $z$-direction, $10$\,{\AA} of vacuum is added to the simulation cell to avoid interactions between periodically translated images. (See Figure \ref{fig:1}(a).) For the 76-atom GB and 108-atom bulk cells, in line with previous studies~\cite{nazarov2012vacancy,nazarov2014ab,fotopoulos2023thermodynamic}, converged 5$\times$4$\times$1 and 4$\times$4$\times$4 $k$-point grids were used, respectively, with a 450\,eV energy cutoff. To determine the number of $k$-points for various cell sizes, the product of the length of the cell and number of $k$-points in the $x$-, $y$-, and $z$-directions was chosen to be as close as possible to $35$ $k$-points $\times$ {\AA} to ensure a converged $k$-point sampling for all unit cells of different dimensions, given the Cu lattice constant of $3.62$\,{\AA}). In the case of the GB cell, due to the added vacuum, just one $k$-point is used in the $z$-direction. The atomic positions were relaxed using energy minimization to an energy tolerance of $<10^{-5}$\,eV.

\subsection{Molecular Dynamics Simulations of Cu-Ti}

All MD simulations were performed using the large-scale atomic molecular massively parallel simulator (LAMMPS)~\cite{thompson2022lammps}. 120,000-atom bicrystalline simulation cells of (210)[100] $\Sigma5$ GBs were constructed using the Atomsk code~\cite{hirel2015atomsk}. For polycrystalline systems, we consider simulation cells with volumes $8\times10^6$~\AA$^3$ containing randomly oriented grains with 665,500 atoms. The polycrystals are then built using radical Voronoi tessellation of randomly distributed points or nodes~\cite{hirel2015atomsk}. Radical Voronoi tessellation allows us to construct complex grain structures, such as curved grain boundaries commonly found in experimental images of polycrystalline metals~\cite{bourne2020laguerre,mantisi2016generation}. Typically 12 nodes are introduced randomly within the periodic cells. The grain sizes obtained from Voronoi tessellation follow a Gaussian distribution with a mean grain volume of 6.6$\times$10$^5$\,{\AA}$^3$ (with a standard deviation of 0.1$\times$10$^6$\,{\AA}$^3$) and mean grain size (MGS) of approximately 89\,\AA. The MGS values are computed using the mean linear intercept (MLI) method~\cite{abrams1971grain}. This value for the MGS is in line with the grain sizes of experimentally reported nanocrystalline Cu~\cite{cao2019bulk,bober2019pronounced} and Cu-based alloys~\cite{chad2020stable}.

Table \ref{tab:comparison} summarizes the MGS, temperature (T), strain rate, and Young's modulus used in the MD simulations, together with previous theoretical and experimental studies of polycrystalline Cu. Table \ref{tab:comparison} illustrates that the elastic moduli obtained from the EAM and MEAM MD simulations are close to the respective values reported in the experimental studies. The cell with an MGS of 8.9\,nm gives an elastic modulus range of 95.7-145.9\,GPa and 92.5-125.3\,GPa using EAM and MEAM MD simulations, respectively. The latter ranges are in good agreement with the reported experimental range of 108-116\,GPa~\cite{sanders1997elastic}, where the samples had considerably larger MGS of 54\,nm.

\begin{table}
\footnotesize

\caption {\label{tab:table1} Mean grain size (MGS) and Young's moduli of polycrystalline pure Cu obtained from previous theoretical and experimental studies at various strain rates and temperatures (T). EMT indicates the many-body effective medium potential approach. For our results, both MEAM and EAM MD simulations are included. Values with $^*$ are obtained using cells with dimensions of one unit cell (one translation, planar cells) along the $z$-axis.}

\begin{tabular}{llllllllll}
   % \hline\hline

 \multicolumn{5}{c}{}\\
   \\  \hline\hline
\cline{1-6}
        & Method & Reference & MGS\,(nm) & T\,(K) & Strain Rate\,(s$^{-1}$) & Young's modulus\,(GPa) &
        \\  \hline\hline
 & EAM & Chen \cite{chen2019temperature} & 4.65-12.41 & 300 & 5$\times$10$^8$ & 54-92 &\\
 & EAM & Zhou \cite{zhou2014effects} & 2.6-53.1 & 300 & 10$^8$ & 25-75$^*$ &\\
  & EAM & Rida \cite{rida2017study} & 9-24 & 300 & 10$^8$ & 55-83 &\\
  & EAM & Xiang \cite{xiang2013molecular} & 2.9-12.6 & 300 & 6.7$\times$10$^7$ & 60-112 &\\
      & EMT & Schiotz \cite{schiotz1999atomic} & 3.28-6.56 & 0 & 5$\times$10$^8$ & 90-120 &\\
           \hline
        & Experimental & Sanders \cite{sanders1997elastic} & 54 & 293 & 10$^{-4}$ & 108-116 &\\
         & & Cheng \cite{cheng2005tensile} & 54 & 297.3 & 10$^{-4}$  & - &\\
     &  & Guduru \cite{guduru2007mechanical} & 23-74 & 293 & 4$\times$10$^{-4}$  & - &
     \\  \hline
        & EAM & (Our results) & 4.1 & 300 & 10$^8$ & 37.2-67.3
 &\\
            & EAM & - & 6.1 & 300 & 10$^8$ & 79.9-104.7
 &\\
            & EAM & - & 8.9 & 300 & 10$^8$ & 95.7-145.9
 &\\
            & EAM & - & 10.2 & 300 & 10$^8$ & 88.8-104
 &\\
           & EAM & - & 12.2 & 300 & 10$^8$ & 93.3-110-5
 &\\
            & EAM & - & 16.3 & 300 & 10$^8$ & 97.7-136
 &\\
              & MEAM & - & 8.9 & 300 & 10$^8$ & 92.5-125.3 & \\
 \hline\hline
\end{tabular}

\label{tab:comparison}
\end{table}

The equilibration of the polycrystalline structures constructed via Voronoi tessellation is crucial for obtaining multigrain simulation cells with structural and mechanical properties comparable to those observed experimentally~\cite{li2019molecular}. Initially, the total energy of all polycrystalline structures is minimized using the conjugate gradient algorithm with a $10^{-10}$\,eV/{\,\AA} tolerance for the atomic forces. To further relax the grain structures, the polycrystals are thermally annealed up to 750\,K at a heating rate of 3\,K/ps for 250\,ps using a time step of 1\,fs at a constant pressure of 1\,bar using the isothermal–isobaric NPT ensemble. The NPT equations of motion are integrated using a leapfrog Verlet algorithm~\cite{kapoor1998determination}. The selected pressure is in line with experimental conditions for Cu-Ti alloys~\cite{chen2015structural}. Previous MD studies in polycrystalline metals, including Cu, have demonstrated that the selected heating rate promotes structural relaxation of the grains without allowing excessive grain growth~\cite{wagih2020learning}. To regulate the temperature, a Nosé-Hoover thermostat~\cite{nose1984molecular,hoover1985canonical} is used with a time constant of 1\,ps, in line with previous MD studies in polycrystalline Cu~\cite{kuppart2021mechanism}. The polycrystals are then cooled to 300\,K at a cooling rate of 3\,K/ps. This cooling rate has been used in previous MD studies for polycrystalline Cu~\cite{wagih2020learning} and Cu alloys~\cite{sargent2021integration}. In the case of the bicrystalline Cu undergoing uniaxial tension, following energy minimization, the system is annealed to 300\,K at a heating rate of 3\,K/ps.

Experimentally, Cu samples containing various concentrations of Ti are prepared in powder form after being mixed in plastic canisters with alumina balls for three hours. Following the mixture process, the powders are sintered at a temperature of 923\,K, with a punch load of 50\,MPa, dwelling time of 5\,min, and a heating rate of 323\,K/min. This process results in Cu-Ti samples with a uniform distribution of Ti~\cite{eze2018effect}. Accordingly, Ti is introduced randomly in polycrystalline simulation cells prior to equilibration. After equilibration, Ti atoms are expected to occupy their most energetically favorable sites. As an example, in the case where 1.5\%at.Ti was randomly introduced into the polycrystalline Cu cell prior to equilibration, 1.4\% of the GB atoms were Ti atoms (approximately 2000 atoms). After equilibration, 1.6\% of the GB atoms were Ti atoms. The 0.2\% increase of Ti atoms in the GBs was due to the motion of atoms near the GBs as the system was annealed. During annealing, no Ti migration was observed from the bulk to the GBs. The latter indicates that the temperature reached during equilibration is not sufficient for Ti to overcome its diffusion barriers in the bulk. The corresponding effective Ti concentrations per grain boundary volume for all simulation cells can be found in Appendix A.

For the bicrystalline and polycrystalline Cu-Ti simulation cells, the uniaxial tensile loading along the $y$-axis is conducted at a strain rate of 10$^8$\,s$^{-1}$, in line with previous MD simulations of Cu~\cite{zhou2017molecular}. During the simulations, the cells are kept at a constant temperature of 300\,K, and the boundaries in the $x$- and $z$-directions are allowed to vary to maintain zero pressure. However, 10$^8$ s$^{-1}$ is a high strain rate compared to those used in most experimental studies. Thus, it may be necessary to extrapolate the stress-strain curves in the simulations to those obtained at much lower strain rates in the experimental systems~\cite{brandl2009strain,ozeren2021prediction}.

\subsection{Energetic Parameters}

\begin{figure}
\centering
\includegraphics[scale=0.035]{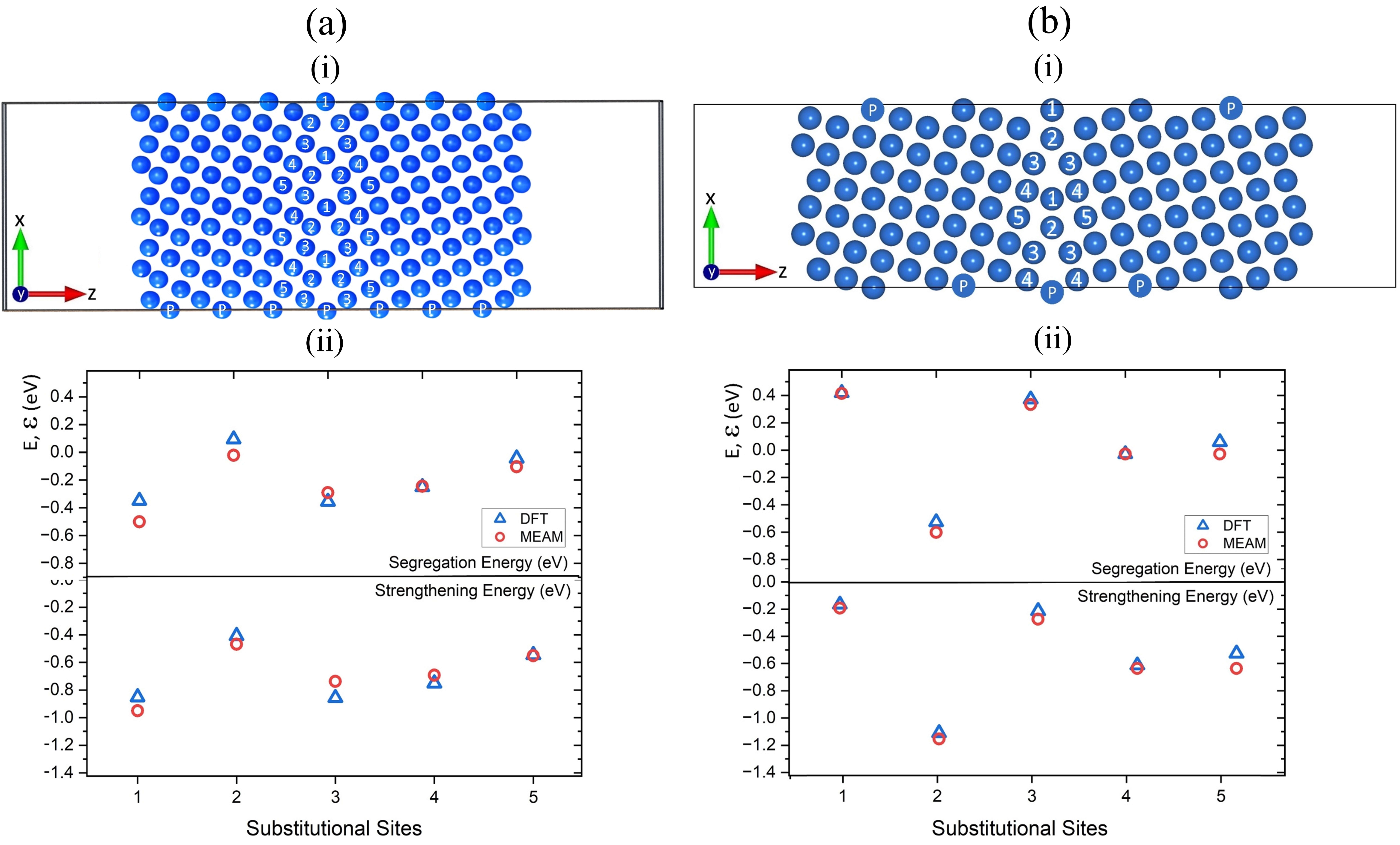}

\caption{\label{fig:epsart} (a) (i) Cu 296-atom (210)[100] $\Sigma5$ GB simulation cell. Cu atoms are shown in blue, and the numbered atoms correspond to the substitutional Ti sites. Atoms labelled with a `P' represent atoms that are periodically translated. (ii) Comparison of Ti segregation (top) and strengthening energies (bottom) obtained using DFT and MEAM for the five substitutional sites. (b) (i) Cu 232-atom (012)[100] GB simulation cell for the labelled segregation sites. (ii) Comparison of Ti segregation (top) and strengthening energies (bottom) from DFT and MEAM.}
\label{fig:1}
\end{figure}

An important component of these studies is the comparison of the results obtained from MEAM MD simulations and DFT calculations. To distinguish between MEAM and DFT energies, $E$ and \begin{large}$\mathcal{E}$\end{large} will indicate the MEAM potential energy and the ground state energy minimized by DFT, respectively. To understand the main properties of Ti in Cu grain boundaries, three energetic parameters need to be computed: the (i) segregation,
(ii) strengthening, and (iii) separation energies. The segregation energies provide information on whether Ti would prefer to be located at the grain boundaries or in the bulk. To calculate the segregation energies of Ti in Cu GB, five substitutional sites are tested, as shown in Figure \ref{fig:1}(a). Since the DFT calculations showed that the octahedral and tetrahedral interstitial sites are unfavorable for Ti, only substitutional sites are considered. The segregation energies of the substitutional Ti sites in Cu grain boundaries are computed using

\begin{equation}
    E/\mathcal{E}_{seg}=(E/\mathcal{E}_{GB+Ti}-E/\mathcal{E}_{GB})-(E/\mathcal{E}_{Bulk +Ti}-E/\mathcal{E}_{Bulk}),
    \label{eq:impurities}
    \end{equation}

\noindent{where} E$/\mathcal{E}_{GB+Ti}$ and E$/\mathcal{E}_{Bulk+Ti}$ are the MEAM/DFT energies of the $76$-atom grain boundaries and $108$-atom bulk Cu cells, respectively, each containing one substitutional Ti atom. E$/\mathcal{E}_{GB}$ and E$/\mathcal{E}_{Bulk}$ are the respective MEAM/DFT pristine Cu grain boundary and bulk energies. Negative energies correspond to favorable segregation of Ti at the GB, whereas positive energies correspond to mixing of Ti in the bulk.

Predictions of the impact of impurities on the grain boundary strength can be made using the strengthening energy, which provides information about whether Ti prefers to be located at the grain boundaries rather than on the sample surface. The MEAM/DFT strengthening energies (E$/\mathcal{E}_{str}$), based on the Rice-Wang \cite{rice1989embrittlement} model, are computed using
\begin{equation}
    E/\mathcal{E}_{str}=(E/\mathcal{E}_{GB+Ti}-E/\mathcal{E}_{GB})-(E/\mathcal{E}_{Sur+Ti}-E/\mathcal{E}_{Sur}),
    \label{eq:strength}
    \end{equation}

\noindent{where} E$/\mathcal{E}_{Sur}$ and E$/\mathcal{E}_{Sur+Ti}$ represent the MEAM/DFT energies of the $108$-atom pure Cu (100) surface and Cu surface simulation cells with one Ti atom, respectively. To obtain the Cu surface simulation cells, a slab approach is employed. Starting from the conventional FCC Cu unit cell, a 3$\times$3$\times$6 translation is performed to create a cell with an extended $z$-dimension. Atoms along the $z$-direction are removed to generate the slab model with a desired slab plus vacuum thickness. We set
the thickness to be 10.86\,{\AA}. A negative value of the strengthening energy means that the impurity will enhance the grain boundary strength, while a positive value implies that the grain boundary will be weakened by the addition of a Ti atom.

Finally, the separation energy indicates whether Ti in Cu GBs promotes or deters fracture initiation. DFT tensile tests use two stretching methods~\cite{lu2006origin,tian2009effect}. In this work, we selected a pre-crack fracture plane based on previous DFT simulations in $\Sigma5$ Cu GBs~\cite{wu2016first,huang2019combined}. The separation energies were calculated by subtracting the total energy of the GB cell at a given spacing between the grains from the energy at the equilibrium separation.

\section{\label{sec:level1}Results}

Before we perform MD simulations of uniaxial tensile strain in bicrystalline and polycrystalline Cu, it is important to compare the results of DFT calculations of GB energetics to those obtained from the MEAM interatomic potential. In this section, we compare the MEAM and DFT energies for one Ti atom introduced into Cu grain boundaries. In addition, we apply uniaxial tension to small CuTi systems and employ DFT to calculate the separation energy for Ti in Cu grain boundaries.

\subsection{\label{sec:citeref}Calculations of segregation and separation energies using DFT and MEAM}

In Figure \ref{fig:1}(a)(i), we show the 296-atom (210)[100] $\Sigma5$ GB simulation cell for five labelled Ti substitutional sites. For the DFT calculations (see Figure \ref{fig:1}(a)(ii)), segregation sites 1 and 3 have the most favorable (lowest) segregation energies, around -0.35\,eV. Previous DFT studies on Al $\Sigma5$ GB solutes showed similar negative segregation energies for Ti, around -0.2\,eV~\cite{karkina2016solute}. In agreement with DFT, MEAM identifies 1 and 3 as the two most favorable Ti substitution sites. For all substitution sites we considered, the MEAM interatomic potential gives segregation energies within 0.15\,eV of those obtained using DFT, which, based on previous comparisons between EAM and DFT calculations~\cite{karkina2016solute,mason2017empirical}, is reasonable agreement between the two methods.

In Figure \ref{fig:1}(a)(ii), we also display the strengthening energies of grain boundaries in the presence of Ti. We used a formalism that is frequently used in the literature~\cite{huang2019combined} to determine the strengthening effect of non-metallic impurities in Cu grain boundaries. We find that MEAM and DFT give negative strengthening energies for Ti in all substitution sites. Thus, both methods predict that Ti would have a strengthening effect when introduced into Cu GBs. The two methods are also in good agreement when a different $\Sigma5$ twin boundary simulation cell is used (i.e. $\Sigma$5 (012)[100] in a simulation cell with 232 atoms as shown in Figure \ref{fig:1}(b)(i)). In both $\Sigma5$ simulation cells, the substitutional site closest to the center of symmetry of the grain boundaries showed the lowest segregation energy (see Figure \ref{fig:1}(b)(ii)), indicating that our results are relevant to other $\Sigma5$ symmetries.

  \begin{figure}
\centering
\includegraphics[scale=0.04]{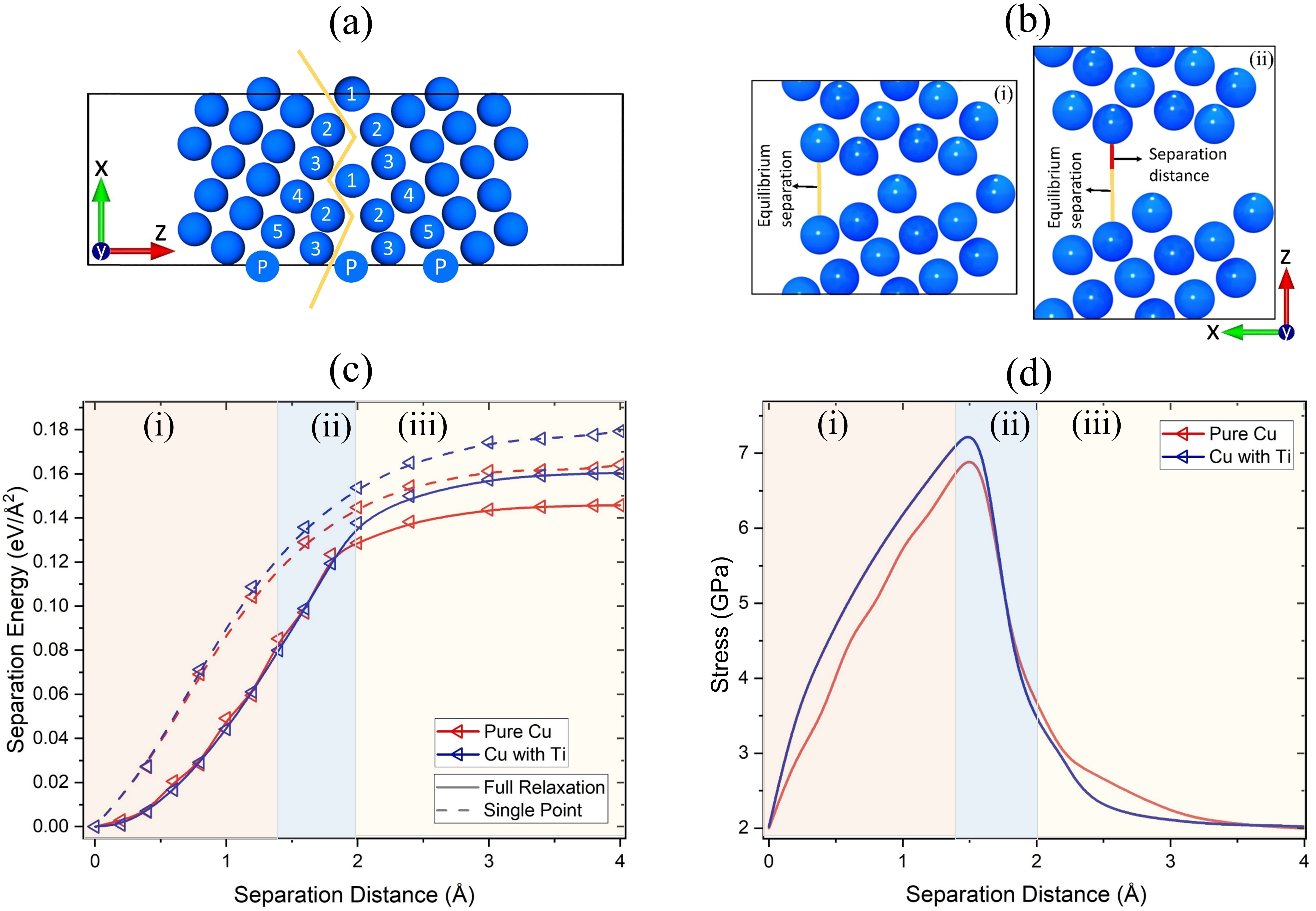}

\caption{\label{fig:epsart} (a) Cu 76-atom (210)[100] $\Sigma5$ GB simulation cell used for the DFT calculations of applied unaxial tension. The yellow line shows the fracture plane. (b) Schematic illustrating the Cu GB (i) at the equilibrium separation and (ii) a separation of 2{\AA}. The yellow and red lines represent the equilibrium and separation distances, respectively. (c) Single point (rigid; no relaxation) and full relaxation DFT calculations of the separation energies, illustrating the energy difference as the separation distance of the grain boundaries is increased for pure Cu and Cu with one substitutional Ti. (d) Hydrostatic stress computed from DFT full-relaxation calculations. The labels (i)-(iii) denote the three distinctive regions during the DFT uniaxial tension calculations.}
\label{fig:2}
\end{figure}

In Figure \ref{fig:2}(a), we show the GB simulation cell used for the DFT uniaxial tension tests. The yellow line illustrates the selected fracture plane. Due to the computational cost of the DFT-based uniaxial tension calculations, we used a smaller cell with 76 atoms. These smaller systems possessed similar segregation and separation energies as those in Figure \ref{fig:1}(a)(i) (see Appendix B). The effect of Ti on the GB separation energy can be seen in Figure \ref{fig:2}(c). The separation distance refers to the displacement along the z-axis from the equilibrium position, as shown in Figure \ref{fig:2}(b). We show the separation energies obtained from both single-point (rigid, no atom relaxation is allowed) and full relaxation DFT calculations. The results follow the universal binding energy relation~\cite{ferrante1981universal}, where the separation energy increases rapidly for small separations and, at larger distances, reaches a plateau. The smallest separations correspond to the pre-fracture regime~\cite{huang2018uncovering} (region (i) in Figures \ref{fig:2}(c) and (d)), while, after $\sim 1$\AA, the system enters the plastic region (region (ii)), where an intergranular fracture initiates. We define a fracture as the point at which uninterrupted areas of zero electron density are formed within the GB. The calculated separation energy in the case of Cu with a single Ti atom is approximately 0.02\,eV/{\AA}$^{2}$ higher than for pure Cu. Another interesting point is that for Cu with a single Ti atom, the separation of the GB into two free surfaces (intergranular fracture) occurs at a separation distance of 2.0\,{\AA}. On the other hand, in pure Cu, the fracture initiates at a separation distance of 1.8{\,\AA}. Finally, in the third deformation stage (region (iii)), the separation energy slowly increases until it plateaus as the remaining long-range interaction forces between the two fracture surfaces tend to zero. The relaxed separation energy curves in the third stage follow the universal binding energy relation.

The tensile strength of the Cu cells with and without Ti can be taken as the maximum tensile stress, shown in Figure \ref{fig:2}(d). The fully relaxed DFT calculations show that the tensile strength of the grain boundary with one substitutional Ti is higher compared to the pure Cu grain boundary. According to these findings, doped Cu showed higher mechanical strength compared to that of pure Cu. The observed effect of Ti is in good agreement with previous DFT studies on dopants in Cu, where it was reported that the introduction of transition metals, such as Nb, Mo, and Zr, into Cu, can significantly increase the energy for the initiation of intergranular fracture~\cite{huang2019combined}. In addition, our results are in good agreement with previous theoretical studies in grain boundaries of Au and Fe, which showed that 3d block transition metals can have a beneficial effect on grain boundary cohesion~\cite{scheiber2021segregation,yamaguchi2019first}.

\begin{figure}
\centering

   \includegraphics[scale=0.045]{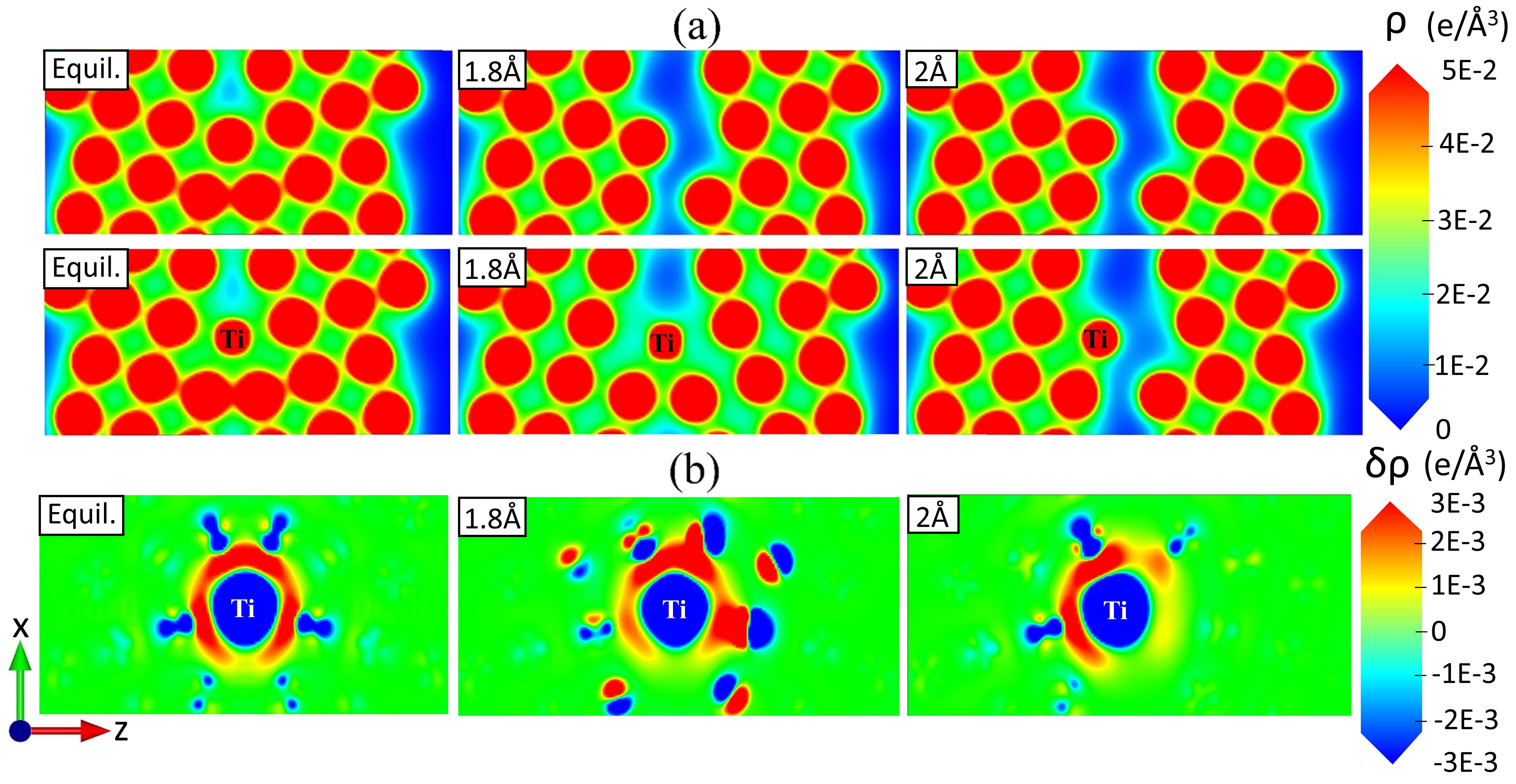}\\

\caption{\label{fig:epsart} (a) Total electron charge density ($\rho$) distribution for a pure Cu GB (top) and a Cu GB with a single Ti atom (bottom) at the equilibrium separation (Equil., first column) and separations of $1.8${\,\AA} (second column) and $2${\,\AA} (third column). (b) 2D contour plots of the differential electron charge density ($\delta$$\rho$) for a Cu GB with a single Ti atom at the equilibrium separation (Equil., first column) and at separations of $1.8$\,{\AA} (second column) and $2$\,{\AA} (third column). Red (blue) shading corresponds to areas of electron accumulation (depletion). We employ VESTA to visualize the electron charge density distributions~\cite{momma2008vesta}.}
\label{fig:3}
\end{figure}

What is the underlying physical mechanism that gives rise to the strengthening of Cu grain boundaries with the addition of Ti? In Figure \ref{fig:3}(a), we show the total charge density distributions ($\rho$) for a pure Cu GB and a Cu GB with one substitutional Ti atom at site 1 (cf. Figure \ref{fig:2}(a)). The three columns indicate the equilibrium GB separation (first column) and separations of $1.8${\,\AA} (second column) and $2${\,\AA} (third column). The three columns also represent the three stages of grain boundary decohesion. At a separation of 1.8{\,\AA}, a fracture initiates in the case of the GB cell without Ti, whereas no fracture is observed in the presence of Ti. The latter effect is attributed to the elongation of the Cu-Cu distance close to the Ti atom. This behavior agrees with the lower separation energy in the case of a Cu GB with Ti as shown in Figure \ref{fig:2}(c).

In Figure \ref{fig:3}(a), we plot the total electron charge density ($\rho$) distribution contours for a pure Cu GB (top) and a Cu GB with a single Ti atom (bottom). In Figure \ref{fig:3}(b), we show the differential charge density ($\delta$$\rho$) contour maps at the equilibrium separation (first column) and at separations of $1.8${\,\AA} (second column) and $2${\,\AA} (third column). Ti, which has two extra valence electrons compared to Cu, significantly redistributes the local charge density at a separation of $1.8${\,\AA}. At this separation, fracture is also initiated in pure Cu, but not in Cu with Ti (see Figure \ref{fig:3}(a)). From the differential charge density distribution contour plots, we find that the addition of Ti enhances the charge density near the GB. As noted in previous studies~\cite{zhu2016first,karkina2016solute}, this charge localization suggests covalent bonding between Ti and the adjacent Cu atoms. Similar covalent bond formation was reported in theoretical studies of Cu-Ti intermetallic compounds, which was attributed to the polarized nature of Ti~\cite{zhu2016first,karkina2016solute}.

In accordance with previous DFT studies of segregants in Cu GBs~\cite{huang2018uncovering}, electron redistribution with strong relaxation reduces the GB strength. However, in cases where relaxation is minimal, such as Cu GBs with Ti substitutions, charge redistribution increases the local strength. Furthermore, in agreement with previous studies~\cite{huang2018uncovering}, dopants in Cu GB that lose electrons are expected to increase the GB strength. Therefore, based on the DFT calculations of Ti in Cu GBs, more energy compared to pure Cu is required to initiate a fracture. Ti enhances the electron density between the two Cu grains, which leads to an increase in the binding energy.

\subsection{\label{sec:citeref}MD Simulations using the MEAM Interatomic Potential}

\subsubsection{\label{sec:citeref}Bicrystalline Cu-Ti under uniaxial tensile strain}

\begin{figure}
\centering
\includegraphics[scale=0.035]{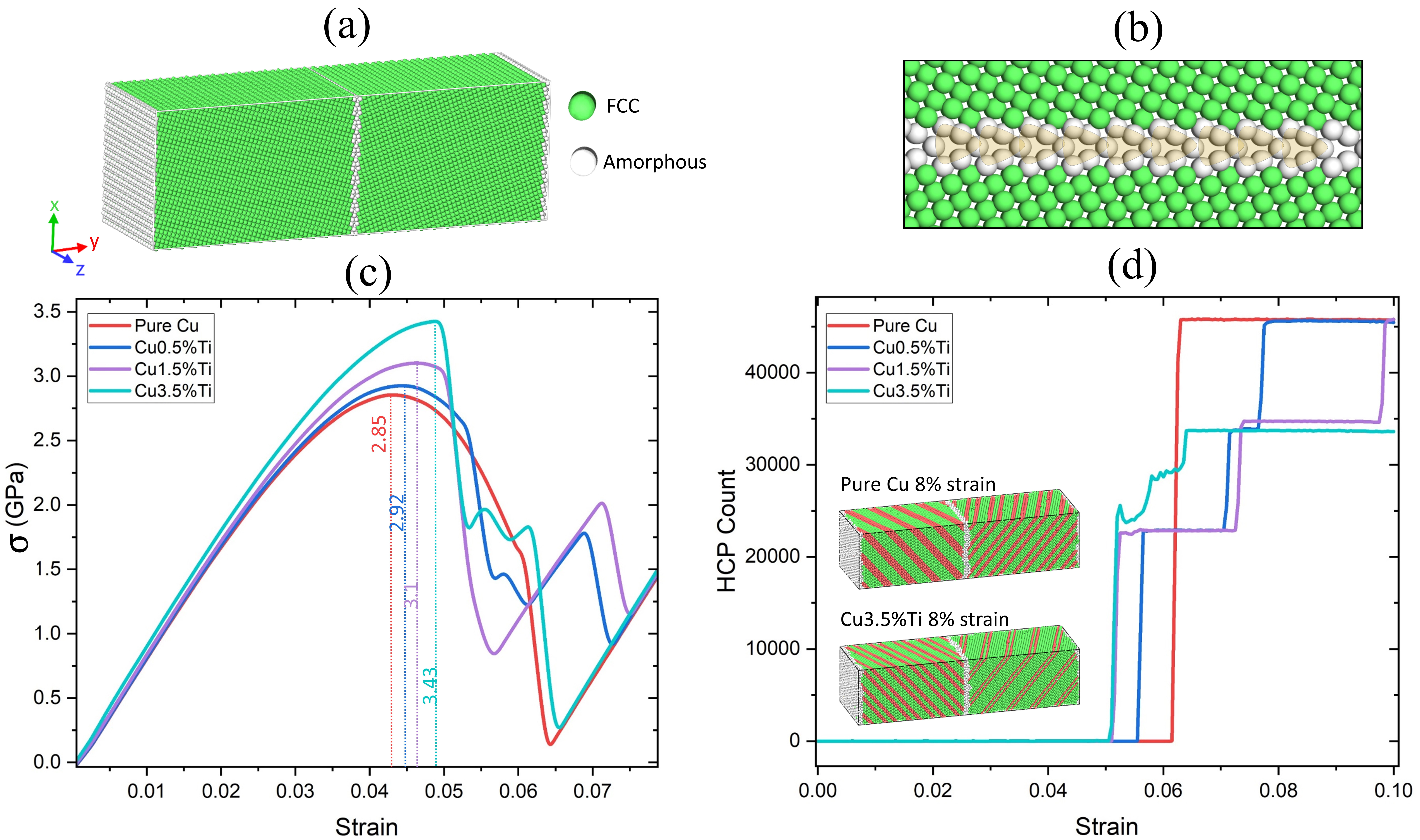}
\caption{\label{fig:epsart} (a) Illustration of the simulation cell used for the MD simulations (using the MEAM interatomic potential) of the bicrystalline  (210)[100] $\Sigma5$ Cu GB (containing 120,000 atoms) undergoing tensile strain. Strain is applied along the $y$-axis. Atoms shaded green and white indicate FCC and amorphous regions of the sample, respectively. (b) GB region of the simulation cell. Yellow shading illustrates the $\Sigma5$ kite-shaped structural units. (c) Stress versus strain plots obtained from the MD simulations as a function of Ti concentration. The vertical lines indicate the yield strength (in GPa) for each concentration. $\sigma$ corresponds to the $yy$ component of the stress tensor. (d) Number of atoms with HCP local order plotted as a function of strain for several Ti concentrations. The insets provide snapshots of the system at 8$\%$ strain for pure Cu (top) and Cu with 3.5$\%$Ti. The red shading indicates atoms with local HCP order. Common neighbor analysis~\cite{faken1994systematic} was used to classify the local crystalline structure surrounding each atom. Structures are visualized using OVITO~\cite{stukowski2009visualization}.}
\label{fig:3.5}
\end{figure}

The results presented in the previous section demonstrate reasonable agreement for the segregation and separation energies of Cu grain boundaries with Ti between the MEAM and DFT methods. We will now carry out MD simulations using the MEAM interatomic potential in larger and more complex Cu GB structures. We will first perform MD simulations of bicrystalline Cu cells (similar to those used in the DFT calculations) undergoing uniaxial tension. The DFT tensile tests revealed that Ti significantly strengthens grain boundaries only for separations $\ge 1.8${\,\AA}. However, these DFT calculations only captured the normal component of the stress, not the shear component, which can induce the formation of topological defects, such as dislocations. By characterizing larger bicrystalline GB cells with MD simulations of uniaxial tensile strain, we can investigate effects of shear stress and dynamics on the yield strength of Cu GBs with varying Ti concentrations.

In Figure \ref{fig:3.5}(a), we illustrate the 120,000-atom simulation cell of a bicrystalline Cu GB with the same symmetry as that used in the DFT calculations of Cu GBs undergoing uniaxial tension (see Figure \ref{fig:3.5}(b)). Constant strain-rate MD MEAM simulations are carried out for pure Cu, as well as Cu with 0.5$\%$, 1.5$\%$, and 3.5$\%$ concentrations of randomly distributed substitutional Ti atoms. Uniaxial tensile tension is applied at constant strain rate after the cell has been equilibrated at a temperature of 300\,K. Figure \ref{fig:3.5}(c) depicts the resulting stress-strain curves, which show that the presence of Ti improves the mechanical properties of the bicrystal, as observed in DFT calculations (see Figure \ref{fig:2}). Both the yield strength (vertical lines in Figure \ref{fig:3.5}(c)) and Young's modulus increase as the Ti concentration increases.

To better understand the underlying mechanism for the strengthening of Cu GBs, we analyzed the local structural order of the bicrystalline GBs as a function of strain and Ti concentration. As the system is strained beyond the yield stress, dislocations are emitted by the GB to release large local concentrations of stress. These dislocations correspond to the formation of stacking faults and twins, which can be identified as planes of atoms with local HCP order. As shown in Figure \ref{fig:3.5}(d), the percentage of HCP atoms decreases with increasing Ti concentration. This result is further illustrated in the insets, where pure Cu possesses a higher density of stacking faults (red-shaded atoms possess HCP order) compared to that for Cu-3.5$\%$ at 8$\%$ strain. Furthermore, Ti is found to hinder the nucleation of the HCP planes in the case of Cu-3.5$\%$Ti, which results in the gradual increase of the number of atoms with HCP symmetry between 5$\%$-6$\%$ strain (see (Figure \ref{fig:3.5}(d))).

\subsubsection{\label{sec:citeref}Polycrystalline Cu-Ti undergoing uniaxial tensile strain}

\begin{figure}
\centering
\includegraphics[scale=0.052]{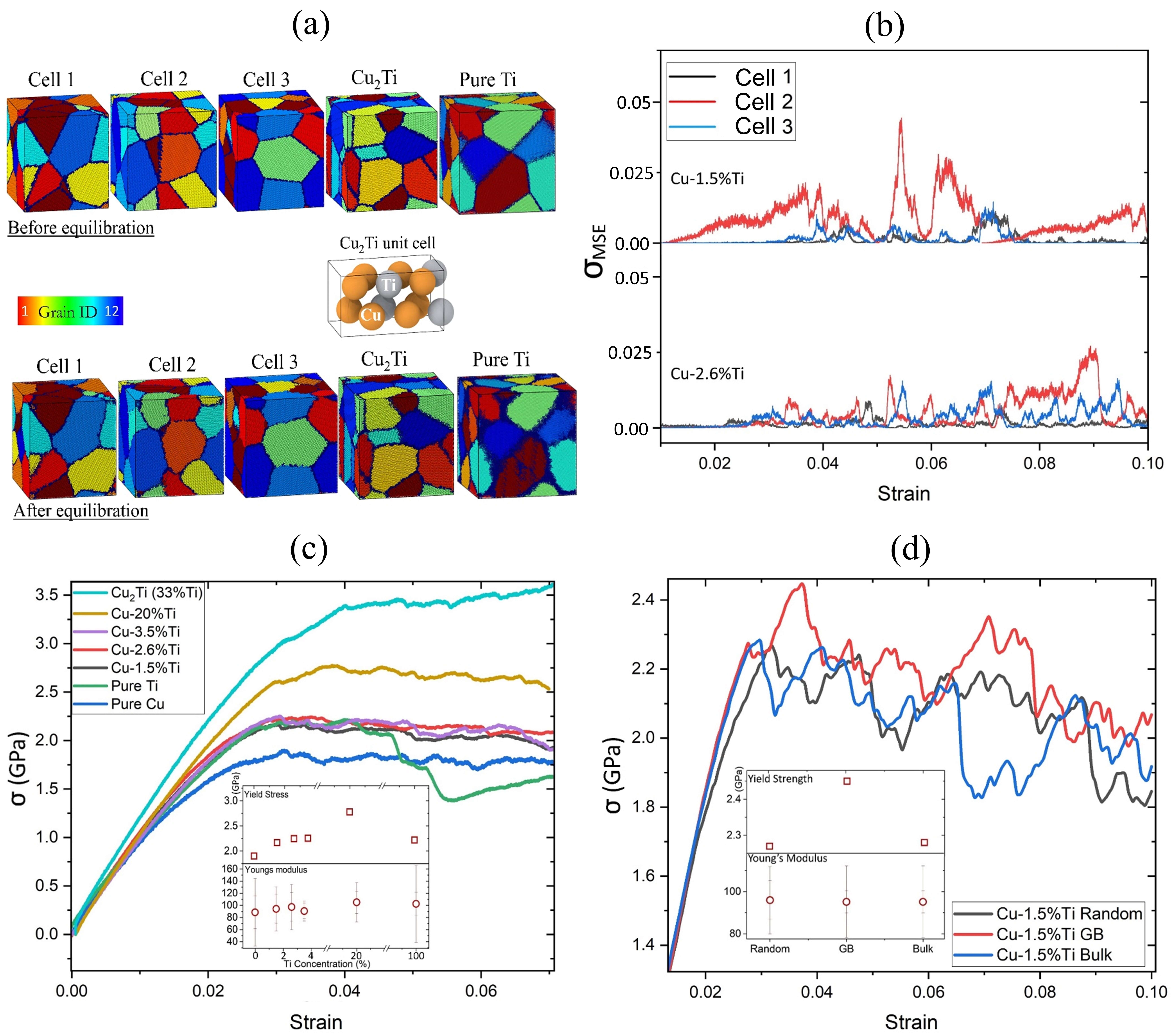}
\caption{\label{fig:epsart} (a) 12-grain polycrystalline cells for Cu-Ti (cells 1-3), Cu$_2$Ti, and pure HCP Ti before (top row) and after (bottom row) equilibration at a temperature of $300$ K. The unit cell of bulk Cu$_2$Ti is also shown, where brown and gray shading indicates the Cu and Ti atoms, respectively. (b) Mean-squared error (MSE) of the stress from the average value for multiple MD MEAM simulations of uniaxial tension with different random initial distributions of Ti in Cu-1.5$\%$Ti and Cu-2.6$\%$Ti using cells 1-3. (c) Stress plotted versus strain for polycrystalline pure Cu, Cu-1.5$\%$Ti, Cu-2.6$\%$Ti, Cu-3.5$\%$Ti, Cu-20$\%$Ti, and Cu-33$\%$Ti (i.e. the Cu$_2$Ti phase). The inset shows the yield strength (top) and Young's modulus (bottom) for several Ti concentrations. (d) Stress versus strain for Cu-1.5$\%$Ti and different initial distributions of Ti: random, only in the GBs, and only in the bulk. The inset shows the yield strength (top) and Young's modulus (bottom) for different initial distributions of Ti. $\sigma$ corresponds to the $yy$ component of the stress tensor.}
\label{fig:4}
\end{figure}

MD simulations of bicrystalline Cu undergoing uniaxial tensile strain showed that Ti hinders the formation of HCP stacking faults, leading to a significant increase in the yield strength. We now examine whether similar strengthening occurs for larger and more realistic polycrystalline models of Cu-Ti. We randomly introduce Ti into the polycrystalline simulation cells with different grain distributions. After adding the Ti atoms, the polycrystals are equilibrated, as described in the Methods section. In Figure \ref{fig:4}(a), we display the polycrystalline Cu-Ti cells prior to (top) and after equilibration (bottom). Cells 1-3 correspond to the cells used for the Cu-Ti polycrystals, with Ti concentrations ranging from 1.5 to 20$\%$. Figure \ref{fig:4}(a) also shows the simulation cells for Cu$_2$Ti and pure HCP Ti.

For each cell and Ti concentration, three MD simulations of uniaxial tension are performed. Different initial random Ti distributions are considered for each of the three runs. In Figure \ref{fig:4}(b), we show the mean-squared error of the stress (MSE; see Appendix A) relative to the average stress over all runs for each of the three simulation cells (cells 1-3 in Figure \ref{fig:4}(a)) for Cu-1.5$\%$Ti and Cu-2.6$\%$Ti. At both concentrations and for cells 1 and 3, the MSE stress is less than 0.02. The MSE in stress continue to decrease with increasing Ti concentration. Interestingly, cell 2 shows the highest MSE in stress for both Ti concentrations. This result can be attributed to the higher fraction of GBs in this cell compared to cells 1 and 3. Even for cell 2, the MSE in stress is $< 0.05$ for Cu-1.5$\%$Ti and $< 0.025$ for Cu-2.6$\%$Ti. Hence, the stress self-averages for different initial random distributions of Ti.

In Figure \ref{fig:4}(c), we show the stress versus strain averaged over multiple runs for pure Cu and Cu-1.5$\%$Ti, Cu-2.6$\%$Ti, Cu-3.5$\%$Ti, and Cu-20$\%$Ti. The 20$\%$ concentration is selected as it corresponds to the point in the Cu-Ti phase diagram where FCC Cu transitions to $\beta$-Cu$_4$Ti~\cite{bateni2003effect}. The stress-strain curve for Cu-33$\%$Ti (Cu$_2$Ti phase), using the cell in Figure \ref{fig:4}(a), is also included as a reference. The inset displays the yield strength and Young's modulus for pure Cu and Cu with several Ti concentrations. In all cases, the inclusion of Ti increases the Young's modulus. Furthermore, the yield strength increases with increasing Ti concentration, in accordance with the DFT calculations and MEAM MD simulations of bicrystalline Cu-Ti undergoing uniaxial tension. Interestingly, for Ti concentrations higher than 1.5\%, the yield strength of Cu-Ti is higher than that of HCP Ti. The mechanical strength of the crystal increases significantly when it transitions into the Cu$_2$Ti phase. This behavior is expected and supported by the solid solution hardening theory for Cu~\cite{vcivzek1974solid}, which shows that doping Cu with metallic substituents will increase its mechanical strength due to the differing mass of the dopant and host atoms~\cite{huang2018uncovering}.

  \begin{figure}
\centering

            \includegraphics[scale=0.053]{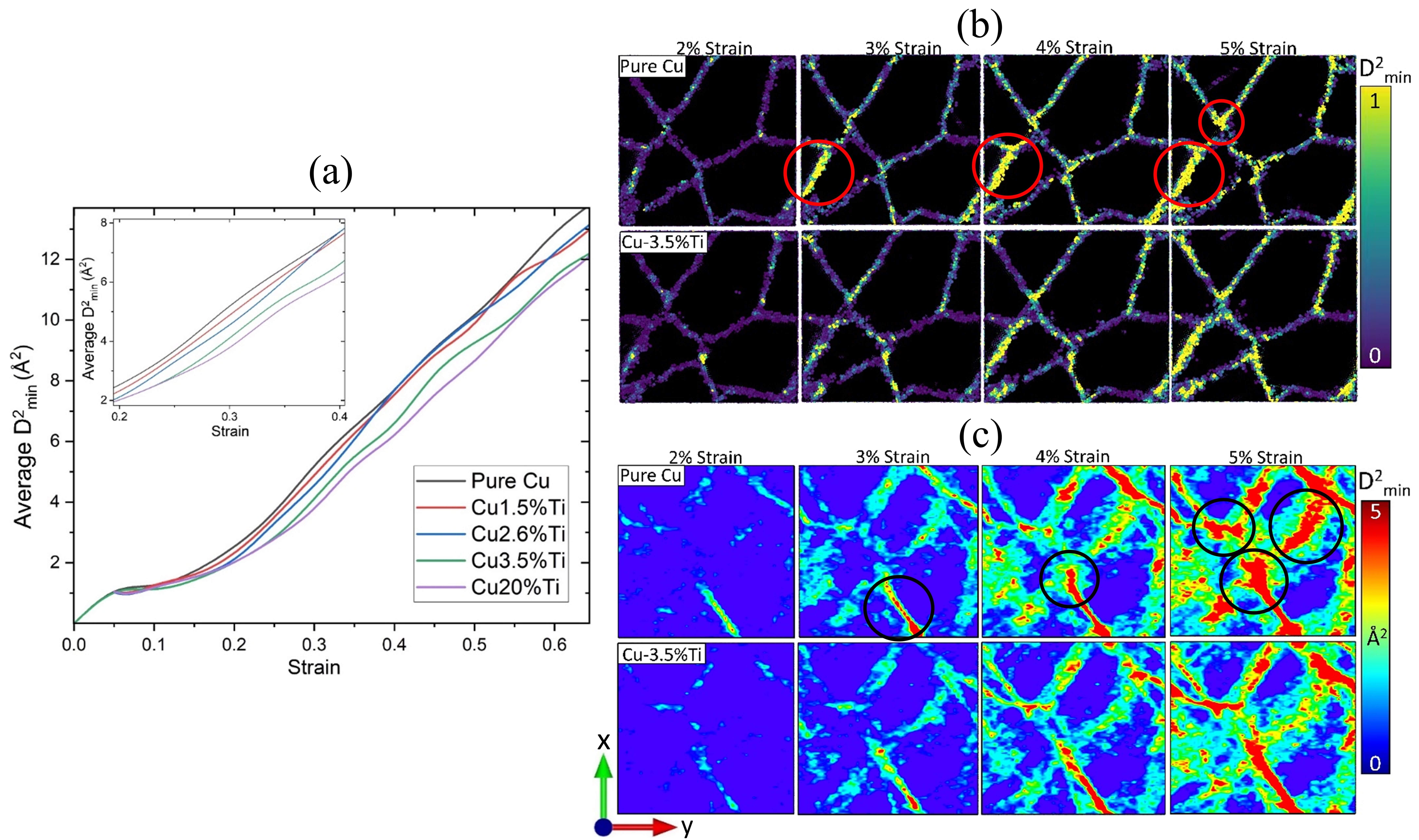}\\

\caption{\label{fig:epsart} (a) Average nonaffine displacement D$^{2}_{\rm min}$ plotted versus strain for pure Cu, Cu-1.5$\%$Ti, Cu-2.6$\%$Ti, Cu-3.5$\%$Ti, and Cu-20$\%$Ti undergoing unaxial tension. Only atoms within the GBs are considered. The inset highlights the region from 2 to 4\% strain. (b) Spatial distributions of D$^{2}_{\rm min}$ at a single $z$-slice ($z=0$) for pure Cu (top) and Cu-3.5$\%$Ti (bottom). (c) D$^{2}$$_{min}$ contour plots averaged for all $z$ for pure Cu (top) and Cu-3.5at.$\%$Ti. For pure Cu, triple junctions surrounding smaller grains possess lower values of D$^2_{\rm min}$ (red and black circles in (b) and (c)).}
\label{fig:5}
\end{figure}

We next seek to determine whether the increase in the yield strength of Cu-Ti compared to pure Cu can be attributed to the presence of Ti in the GBs. In Figure \ref{fig:4}(d), we show the stress versus strain for Cu-1.5$\%$ when Ti was initially randomly distributed, was located only in the bulk, and only in the GBs. In the latter case, Ti is introduced at random sites, but only within the grain boundary region, as defined by common neighbor analysis. As shown in the inset of Figure \ref{fig:4}(d), the inclusion of Ti, either only in the bulk or only in the GBs does not affect the Young's modulus for small strain. However, the stress versus strain curve for randomly distributed Ti has a smaller Young's modulus for strains $>2\%$. Furthermore, when Ti is introduced only in the GB region, we find a considerable increase in the yield strength compared to the case where Ti is distributed randomly or only in the bulk (see inset in Figure \ref{fig:4}(d)). These results illustrate that decorating Cu GBs with Ti atoms will significantly improve the mechanical strength of the crystal.

\begin{figure}
\centering

           \includegraphics[scale=0.055]{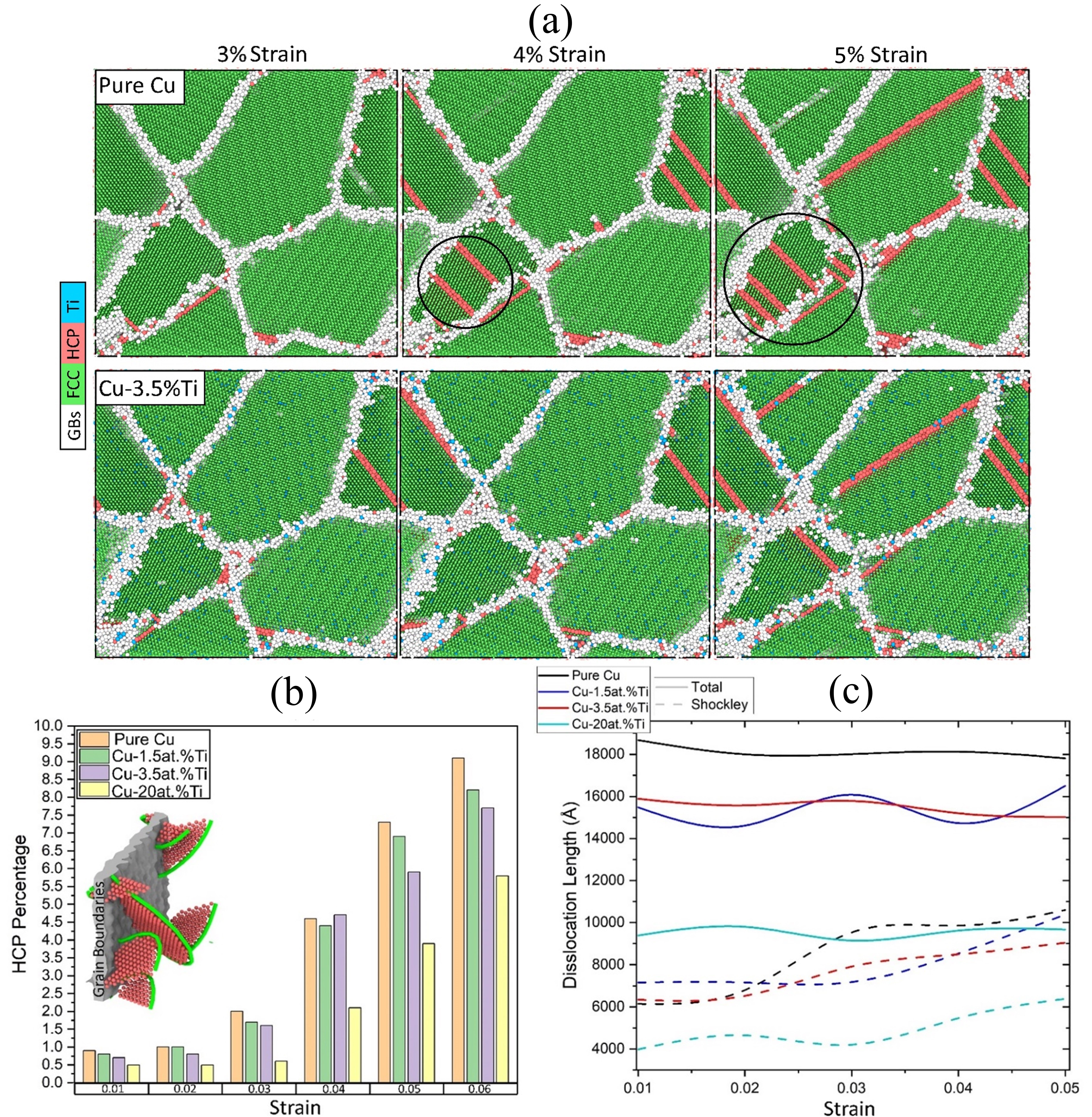}

\caption{\label{fig:epsart} (a) Characterization of the local structural order for (top) pure Cu and (bottom) Cu-3.5\%Ti at three strains. Ti (blue) atoms prevent the formation of HCP order (red shading) in smaller grains and reduce the number of grain-boundary induced dislocations (black circles). (b) Percentages of atoms with HCP order for Cu-3.5\%Ti at several strains. The inset illustrates the initiation of a stacking fault (HCP atoms in red) via the emission of partial Shockley dislocations (green lines) from the GB (gray region). (c) Total length of all dislocations and Shockley dislocations versus strain for pure Cu, Cu-1.5$\%$Ti, Cu-3.5$\%$Ti, and Cu-20$\%$Ti.}
\label{fig:6}
\end{figure}

To better understand the origin of the superior mechanical properties of Cu-Ti, we examine the mechanical response between 2$\%$ and 5$\%$ strain, approximately where pure Cu and all Cu-Ti alloys reach their yield strength (see Figure \ref{fig:4}(c)). (Because polycrystalline cell 1 (see Figure \ref{fig:4}(a)(i)) has the lowest MSE in stress among the three cells tested, this cell is used for the following analysis.) In Figure \ref{fig:5}(a), we show the average non-affine displacement D$^2$$_{\rm min}$ as a function of strain for pure Cu and Cu with increasing Ti concentration. (Only atoms within the GBs are included in D$^2$$_{\rm min}$; see Appendix A for more details.) For strains larger than 2$\%$, cells with a higher concentration of Ti have lower D$^2$$_{min}$ values. The spatial distribution of D$^2$$_{min}$ for pure Cu and Cu-3.5$\%$Ti at strains between 2$\%$-5$\%$ are shown in Figures \ref{fig:5}(b) and (c) for a single $z$-slice and averaged over all $z$, respectively. In the case of pure Cu, the GBs of the smaller grains show higher values of D$^2$$_{\rm min}$ (see the red and black circles in Figures \ref{fig:6}(b) and (c), respectively). The triple junctions and GBs surrounding smaller grains show considerably smaller D$^2$$_{\rm min}$ in Cu-3.5$\%$Ti.

In Figure \ref{fig:6}(a), we show that intrinsic stacking faults (ISFs), which are characterized by atoms with local HCP order (red shading) begin to nucleate from the GBs and propagate at 3$\%$ strain. Consistent with the results from the tensile tests of bicrystalline Cu-Ti, the presence of Ti (blue atoms) prevents the formation of HCP-ordered atoms in smaller grains. Thus, the GBs in Cu-3.5$\%$Ti emit a lower density of partial dislocations compared to pure Cu (see black circles in Figure \ref{fig:6}(a)). Based on these results, the dislocations emitted by the grain boundaries may be responsible for both the higher values of D$^2$$_{\rm min}$ and the lower yield strength of pure Cu compared to the Cu-Ti alloys.

Figure \ref{fig:5}(b) shows the fraction of atoms with local HCP order for Cu-Ti alloys up to 6$\%$ strain. Higher Ti concentrations reduce the number of atoms with HCP order. This effect aligns with what was observed in the MD simulations of bicrystalline Cu-Ti undergoing uniaxial tension. Pure Cu has more HCP atoms at 5$\%$ strain than
Cu-20$\%$Ti has at 6$\%$ strain. The inset in Figure \ref{fig:5}(b) illustrates the initiation of a stacking fault (HCP atoms in red) via the emission of Shockley dislocations by the grain boundaries. We find that Ti inhibits the emission of Shockley dislocations and nucleation of stacking faults. This dislocation inhibition phenomenon will become more prominent when Ti is introduced in the GBs rather than randomly. The latter explains the enhanced mechanical properties of Cu-Ti under strain when Ti was introduced only in the GBs (see Figure \ref{fig:6}(d)).

Since dislocations govern the mechanical properties of polycrystals~\cite{li2020emission,karkina2016solute}, we now examine and quantify the effect of Ti on the dislocation emission process. In Figure \ref{fig:6}(c), we show the lengths of all dislocations in pure Cu, Cu-1.5$\%$Ti, Cu-3.5$\%$Ti, and Cu-20$\%$Ti as a function of strain. Since partial dislocations (stacking faults) are initially emitted as Shockley dislocations by the GBs~\cite{hu2022formation}, the lengths of the Shockley dislocations are also included for each concentration. Increasing the Ti concentration increases the lengths of both the total and Shockley dislocations. In pure Cu, Shockley dislocations have a larger total length than that for Cu-20$\%$Ti at 3$\%$ strain. Increasing the Ti concentration decreases both the total and Shockley dislocation lengths. Thus, Ti is expected to reduce twinning, a process during which a part of the metal undergoes shear deformation along a preferred slip plane. The results in Figures \ref{fig:6}(a) and (b) are consistent with twinning, where the addition of Ti is found to alter the crystal's ability to generate HCP planes. A similar finding was reported in bicrystalline Al through MD simulations, where the inclusion of Mg in the GB hindered the nucleation of dislocations, leading to an increased yield strength~\cite{ma2022unraveling}.

Based on our analysis and taking into account the already established role of dislocation emission in nanocrystalline FCC metals~\cite{li2020emission}, we can identify a mechanism where the inclusion of Ti atoms improves the mechanical properties of the samples under tensile loading. Initial plastic deformation occurs under tensile loading throught the nucleation of partial dislocations emitted by the GBs at a strain of $\sim 1.5\%$. The continual process of partial dislocation nucleation from GBs and annihilation of GBs at opposite ends of the grains causes pronounced local distortion close to the GBs. With the addition of Ti, the distortion becomes less pronounced. Ti, when introduced at the GBs, increases the energy needed for the GBs to slide and emit partial dislocations. Thus, the inclusion of Ti substantially decreases the density of the emitted partial dislocations by the GBs, leading to an increase in the yield stress. The latter observation is in agreement with the Hall–Petch equation~\cite{hall1951deformation,petch1953cleavage}, which predicts that as the stress needed for the activation of the dislocations increases, the yield strength also increases.

\section{\label{sec:level1}Summary and Conclusions}

Previous studies have mainly used DFT calculations to investigate the role of metallic solutes in the grain boundaries of Cu in determining its mechanical properties. To better understand the effect of Ti on the mechanical properties of Cu, we employ both DFT calculations and MEAM MD simulations, which enables studies of tensile deformation over a range of length scales. We first compared the results from the MEAM potential for the most energetically favorable Ti substitution sites in Cu GBs to the results from the DFT calculations. For both the DFT calculations and MEAM MD simulations, substitutional Ti prefers to segregate at the GB rather than in the bulk of Cu. DFT calculations of tensile deformation show that Ti at the GBs induces local charge localization, which increases the maximum stress, and the separation and strengthening energies of the GB. An interesting future direction would be to examine whether other elements with properties similar to Ti, like the valency of electrons or radius, when added to Cu grain boundaries would cause similar charge localization and increases in the maximum stress and segregation and separation energies.

The changes in the mechanical properties of large bicrystalline Cu cells in response to the addition of Ti was studied using MEAM MD simulations of uniaxial tension. Similar to the results obtained from DFT calculations of uniaxial tension, the addition of Ti increased the yield strength of the crystal, due to the inhibition of stacking faults emitted by the GBs. This phenomenon was also present when polycrystalline systems were considered. MEAM MD simulations using nanocrystalline multigrain simulation cells showed that the presence of Ti prevents the emission of partial Shockley dislocations from GBs, which reduces local distortions. This effect becomes more pronounced as the Ti concentration increases and when all Ti atoms are introduced at the GBs of the polycrystal. Thus, the inclusion of Ti results in a significant increase in the yield strength and Young's modulus of Cu polycrystals compared to that for pure Cu polycrystals. These mechanical properties were further improved at higher Ti concentrations.

Our findings suggest that the addition of Ti to Cu has a significant effect on the yield strength and Young's modulus of the material. Thus, grain boundary segregation engineering of polycrystalline Cu with Ti solutes can be used to enhance the material's durability and broaden its industrial applications. However, due to the intrinsic limitations of the time and length scales of MD simulations, the unrealistically high deformation rates used in MD simulations are many orders of magnitude higher than those used in experiments. Alongside the current data, future work should focus on extrapolating the mechanical properties of Cu-Ti to significantly slower strain rates. In addition, further investigations are needed for a broader range of Cu-Ti alloys. Finally, to simulate polycrystalline Cu-Ti cells with mean grain sizes closer to those studied experimentally, reliable and less computationally demanding MEAM potentials for modeling polycrystalline Cu-Ti will be required.

\newpage

\begin{suppinfo}
\end{suppinfo}


%% classes.  This should be given within the "acknowledgment"
\begin{acknowledgement}

A.L.S. acknowledges funding by EPSRC (grant EP/P013503/1). C.S.O. acknowledges funding from NSF Grant No. 2244310. V.F. would like to acknowledge funding by EPSRC (grant EP/L015862/1) as part of the CDT in molecular modeling and materials science. V.F. gratefully acknowledges the UCL Doctoral School for supporting the placement in the UCL-Yale Collaborative Exchange Program. Computational resources on ARCHER2 (http://www.archer2.ac.uk) were provided via our membership of the UK's HPC Materials Chemistry Consortium, which is funded by EPSRC (EP/L000202, EP/R029431).

\end{acknowledgement}

\bibliography{Manuscript}

\end{document}